\documentclass[aps,prl,reprint,floatfix, superscriptaddress]{revtex4-1}

\usepackage[pdftex]{graphicx,color}
\DeclareGraphicsExtensions{.pdf,.png}

\usepackage{grffile}

\usepackage{amsmath, amsthm, amssymb, amsfonts, amsbsy}
\usepackage{mathtools} 
\usepackage{bm}
\usepackage{mathrsfs}  

\definecolor{strawberry}{rgb}{1.0,0.0,0.5}

\definecolor{blueberry}{rgb}{0.015686275,0.2,1.0}

\usepackage[capitalise]{cleveref}
\crefformat{equation}{Eq. (#2#1#3)} 
\Crefformat{equation}{Equation (#2#1#3)} 
\crefrangeformat{equation}{Eqs. (#3#1#4--#5\crefstripprefix{#1}{#2}#6)} 
\Crefrangeformat{equation}{Equations (#3#1#4--#5\crefstripprefix{#1}{#2}#6)} 
\crefmultiformat{equation}{Eqs. (#2#1#3)}{ and~(#2#1#3)}{, (#2#1#3)}{ and~(#2#1#3)} 
\Crefmultiformat{equation}{Equations (#2#1#3)}{ and~(#2#1#3)}{, (#2#1#3)}{ and~(#2#1#3)} 

\usepackage{xcolor}

\begin{document}

\title{Traveling strings of active dipolar colloids}

\author{Xichen Chao}
\affiliation{School of Mathematics, University of Bristol - Bristol BS8 1UG, UK}
\author{Katherine Skipper}
\affiliation{H.H. Wills Physics Laboratory, Tyndall Avenue, Bristol, BS8 1TL, UK}
\author{C. Patrick Royall}
\affiliation{H.H. Wills Physics Laboratory, Tyndall Avenue, Bristol, BS8 1TL, UK}
\affiliation{Gulliver UMR CNRS 7083, ESPCI Paris, Universit´e PSL, 75005 Paris, France}
\author{Silke Henkes}
\affiliation{School of Mathematics, University of Bristol - Bristol BS8 1UG, UK}
\affiliation{Lorentz Institute, LION, Leiden University - Leiden 2333 CA, NL}
\author{Tanniemola B. Liverpool}
\affiliation{School of Mathematics, University of Bristol - Bristol BS8 1UG, UK}

\date{\today}

\begin{abstract} 
We study an intriguing new type of self-assembled active colloidal polymer system in 3D. It is obtained from a suspension of Janus particles in an electric field that induces parallel dipoles in the particles as well as self-propulsion in the plane perpendicular to the field. At low packing fractions, in experiment, the particles self-assemble into 3D 
columns that are self-propelled in 2D. Explicit numerical simulations combining dipolar interactions and active self-propulsion find an activity dependent transition to a string phase by increasing dipole strength. We classify the collective dynamics of strings as a function of rotational and translational diffusion. Using an anisotropic version of the Rouse model of polymers with active driving,  we analytically compute the strings' collective dynamics and centre of mass motion, which matches simulations and is consistent with experimental data.  We also discover long range correlations of the fluctuations along the string contour that grow with the active persistence time, a purely active effect that disappears in the thermal limit.
\end{abstract}

\maketitle

Active matter describes a new class of materials that are composed of elements driven out of equilibrium by internal sources of energy. 
{These systems promise a novel way to add}
functionality in materials design for a variety of applications, from drug delivery to metamaterials~\cite{baylis2015self,di2010bacterial, krishnamurthy2016micrometre,stenhammar2016light, frangipane2018dynamic, arlt2018painting,chen2021realization}.
A  major challenge however is how to {\em control} activity, i.e.  which components of a system are active, when that activity is to be switched on/off, and how to use it to steer emergent collective behaviour towards a desired function. 
One promising avenue for controlling active matter is by tuning the interplay between active driving and passive mechanics of 
the mesoscale structures of the active material at scales intermediate between the microscopic building blocks and the macroscopic (hydrodynamic) scales~\cite{marchetti2013hydrodynamics}.
This dynamic structure at the mesoscale can take the form of polymers~\cite{humphrey2002active,liverpool2001viscoelasticity,eisenstecken2016conformational,chakrabarti2020flexible}, membranes~\cite{al2020dynamics}, and disordered or ordered solids~\cite{henkes2020dense,caprini2020spontaneous,tan2022odd,scheibner2020odd}. Because of the complex internal dynamics of these meso-structures, more detailed descriptions, going beyond long wavelength hydrodynamics, must be developed to precisely uncover the physical principles required to accurately control their behaviour. 

Extended one-dimensional polymeric structures are a promising mesoscale ingredient as their 
 relatively open structure leaves the system more susceptible to external controls. 
Hence there has been a corresponding resurgence of experimental and theoretical  work on active polymer systems~\cite{schaller2010polar,deblais2023worm,fazelzadeh2023effects}. Most experimental realisations of active polymer systems have been biological, e.g. motor-driven cytoskeletal polymers~\cite{mizuno2007nonequilibrium} or living organisms such as worms~\cite{deblais2023worm}. Theoretical studies have included 
tangentially driven linear polymers and ring polymers, mostly in 2D~\cite{Liverpool2003a,isele2015self, prathyusha2018dynamically,fazelzadeh2023effects} and more recently have begun to look at entanglement~ \cite{mandal2020crowding}. 
Biological components however are hard to control and there is a need for systems built from man-made (synthetic)
components.

Active Janus colloids are one of the simplest experimental building blocks of synthetic active materials -  their single particle dynamics is reasonably well approximated by active Brownian particles \cite{marchetti2016minimal,bechinger2016active,cates2015motility}. 
{The Janus particles of interest here} are made from an insulating colloid half coated by metal {which is itself then coated in a layer of insulator}. When placed in an oscillating electric field, by the process of induced-charge electrophoresis (ICEO)~\cite{squires2006breaking, gangwal2008induced}, they 
simultaneously {become} active and interact via pairwise dipolar interactions. 
The colloids undergo sedimentation to the bottom of the sample. Hence they can self-assemble on the bottom surface  into 2D  polymer-like motile chains~\cite{patteson2016active,yan2016reconfiguring, zhang2021active}. 
 Recent experiments have shown however that it is possible to study this system in fully 3D by suspending smaller Janus particles in a solvent, which due to their size sediment markedly less~\cite{sakai2020active}. 
 At low volume fractions, the particles self-assemble into active columns (strings) that self-propel in the plane perpendicular to their axis.  We note 
 it is well known from 
 experiments \cite{tao2001super,Yethiraj2003} 
 and simulations \cite{hynninen2005phase} that a passive collection  of suspended dipolar colloids have a {\em static} string phase at low density.

\begin{figure*}[htbp]
    \centering
    \includegraphics[width=\linewidth]{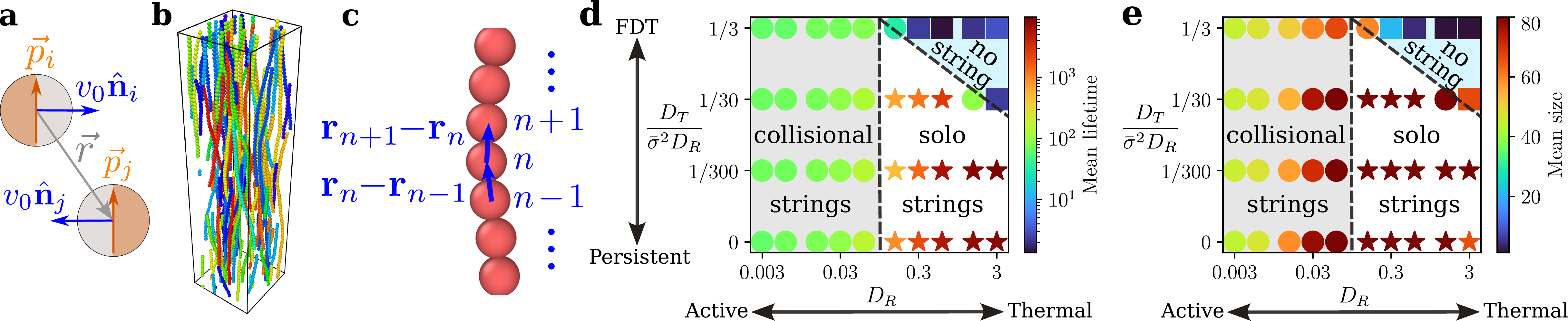}
    \caption{\textbf{a} - Schematic of the Janus particle used in experiments. The dipoles $\mathbf{p}_i=(0,0,p)$ are aligned with electric field direction and perpendicular to self-propelling velocities $\mathbf{v}_0=v_0 (\cos \theta_i,\sin \theta_i,0)$. $\mathbf{r}=(x,y,z)$ is the vector separating the positions of two bead centres. \textbf{b} - Simulations conducted in LAMMPS and visualised in OVITO \cite{stukowski2009visualization}, with packing fraction $5\%$. Different colors indicate different clusters. \textbf{c} - The bond vector for a single string, with $\bm{r}_n$ the position of the $n$th bead. \textbf{d-e}, Phase diagrams of the mean lifetime and mean size of traveling strings. $p=1, v_0=0.5$ as a function of rotational diffusion coefficent $D_R$ and ratio $\frac{D_T}{\Bar{\sigma}^2D_R}$. The three regimes and markers in these phase diagrams are determined by the mean lifetime phase diagram. 
    The colourbar in the mean size phase diagram is linear and is logarithmic in the mean lifetime phase diagram. These phase diagrams are based on simulations with box size $20\times20\times80$. 
    }
    \label{fig:introductory}
\end{figure*}

In this letter, we study the collective dynamics of the low density phase of actively travelling strings through a combination of numerical simulations, theoretical tools, and experimental analysis. Our numerical model combines short range repulsion,  dipolar interactions {in the direction of the field} and self propulsion, and we find an activity-dependent transition to the active string phase as a function of dipole strength. We explore the collective dynamics of strings as a function of rotational and translational diffusion strengths. Using a generalised Rouse model of anistropic 3-dimensional flexible polymers with active driving allows us to capture the string dynamics. In addition to explaining the global string dynamics, this predicts a purely active emergent correlation length along the strings. We further carry out experiments with metallo-silica Janus particles that we study with confocal microscopy at the single-particle level~\cite{ivlev2012complex}. We verify that our theoretical model agrees very well with simulations and is consistent with experiments.


\textit{Simulation ---}  We model our dipolar active colloids by adopting a hybrid potential that combines the Weeks-Chandler-Anderson (WCA) potential for short-range repulsive interactions between particles with a dipole-dipole (DD) pair interaction that accounts for the long-range dipolar forces present in the system \cite{hynninen2005phase}.
In experiment, the dipole moments are aligned with the oscillating electric field $\mathbf{E} = E\hat{\bm z}$~\cite{yan2016reconfiguring,sakai2020active}. Hence all particles have a constant identical dipole moment $\mathbf{p} = p \hat{\bm z}$ with dipole strength $p$ (that increases with $|E|$). 
The interaction potential between particles $i$ and $j$, 
is  
%
$U_{HY, ij}=U_{\mbox{\tiny WCA}}(r_{ij})+U_{\mbox{\tiny DD}}(\mathbf{p},r_{ij})$ with 
\begin{align}
\begin{split}
    U_{\mbox{\tiny WCA}} (r) &=4\epsilon\left[\left(\frac{\sigma}{r} \right)^{12}- \left(\frac{\sigma}{r} \right)^{6}\right], ~\ r<2^{1/6}\sigma, \\
    U_{\mbox{\tiny DD}}(\mathbf{p},r)&=\frac{1}{r^3}(\mathbf{p}\cdot \mathbf{p})-\frac{3}{r^5}
    (\mathbf{p}\cdot \mathbf{r})(\mathbf{p}\cdot \mathbf{r}), \label{potential}
\end{split}
\end{align}
where $\mathbf{r}_{ij} = \mathbf{r}_{i} - \mathbf{r}_j$ is the inter-particle separation and $r_{ij} = |\mathbf{r}_{ij}|$. We choose units such that the WCA potential strength $\epsilon=1$ and the particle {diameter} is $\bar{\sigma} = 2^{1/6} \sigma = 1$. The DD pair interaction is attractive along the direction of the polarisation ${\mathbf{p}}$ and repulsive in the orthogonal plane, naturally leading to the formation of parallel strings along $\hat{\bm z}$. Due to ICEO, each Janus particle rotates until the interface between the two halves is parallel to $\mathbf{E}$, and therefore self-propels in a direction in the $xy$ plane (Fig. \ref{fig:introductory}a).   The Debye screening length in experiment is much smaller than particle diameter, allowing us to cut off dipole interactions after the first neighbour for computational efficiency.

We combine $U_{HY, ij}$ with overdamped Active Brownian dynamics without hydrodynamics,
\begin{align}
\begin{split}
     \dot{\mathbf{r}}_i&=-\frac{1}{\zeta}\sum_{j\neq i}\pmb{\nabla }_i U_{HY,ij}+v_0\hat{\mathbf{n}}_i+\pmb{\eta}_i^T,\\
    \hat{\mathbf{n}}_i&=(\cos \theta_i, \sin \theta_i, 0), ~\ \dot{\theta}_i=\eta_i^R,
\end{split}
\end{align}
where $\{\mathbf{r}_i, \theta_i \}$ are the position, orientation of the $i$th particle and  $\zeta$ is the Stokes drag.
{\color{black}{We include activity with self-propulsion speed $v_0$ (that in experiment increases with $|E|$) in the direction $\hat{\mathbf{n}}_i$, which is constrained to the $xy$ plane, orthogonal to $\hat{\bm z}$ }}
%
(Fig. \ref{fig:introductory}a). Its in-plane angle $\theta_i$ diffuses with rotational white noise $\eta^R_i$, with correlation $\langle \eta_n^R(t)\eta_m^R(t') \rangle=2D_R \delta(t-t')\delta_{nm}$, where $D_R$ is the rotational diffusion coefficient.
We also include translational white noise $\pmb{\eta}_i^T$ in all directions, with correlations $\langle \eta_{\alpha n}^T(t)\eta_{\beta m}^T(t') \rangle=2D_T \delta(t-t')\delta_{\alpha \beta}\delta_{nm}$ where $D_T$ is the translational diffusion coefficient.
The simulation box is periodic in all three dimensions and we use LAMMPS~\cite{thompson2022lammps} with a custom ABP integrator~\cite{cameron2023equation}. 

In our simulations, we systematically vary the dipole strength, speed and diffusion coefficients via the parameters $(p, v_0, D_R, {\frac{D_T}{\Bar{\sigma}^2D_R}})$ while maintaining a fixed low packing fraction $\phi=0.05$. Strings are defined by a clustering algorithm  with neighbour cutoff distance $1.05\bar{\sigma}$ (Fig. \ref{fig:introductory}b). 
Here we first locate the transition to the string phase by varying $p$ and $v_0$ independently for intermediate $D_R =0.15$ and different $D_T$ (Fig. S2.1a-d). The transition from a disordered phase at low $p$ to a string phase at high $p$ shifts from around $p=0.2-0.3$ at $v_0=0.1$ to $p\lesssim 1$ at $v_0=0.7$. We choose the point $p=1$ and $v_0=0.5$, which is in the string phase in almost all cases. We find that string formation is subject to slow coarsening dynamics, necessitating runs of $t=20000$ time units to reach steady-state (Fig. S2.2a). Strings also rapidly lengthen when $p$ increases (Fig. S2.2b), and we switch to a tall simulation box $L_x \times L_y \times L_z \equiv 20\times20\times80$ for our main runs (Fig. \ref{fig:introductory}b). Strings can still span the system, so we cut off string size at $L_z/\bar{\sigma}=80 $.   

The persistence of active driving is regulated by rotational diffusion, $D_R$  
and translational diffusion, $D_T$. If the fluctuation-dissipation theorem (FDT) holds,  $D_T = k_B T/\zeta$ and $D_R = k_B T/\zeta_r$.  For Stokes drag, where ${\Bar{\sigma}^2\zeta = 3\zeta_r}$  this implies that {$\frac{D_T}{\Bar{\sigma}^2D_R} = \frac{1}{3}$} in simulation units.
 Another limit, considered in many ABP simulations is where orientational noise dominates and one can set $D_T \approx 0$ (``persistent''). Finally when $D_R$ is very large the effect of activity is to give an effectively thermal system with a renormalized {\em active temperature} (``thermal''). The two axes of our phase diagrams are then $D_R$ and ${\frac{D_T}{\Bar{\sigma}^2D_R}}$, with $D_R$ varying from $0.003$ to $3$, which correspond to an active limit and a thermal limit respectively. ${\frac{D_T}{\Bar{\sigma}^2D_R}}$ varies from $0$ to $\frac{1}{3}$, which indicate a persistent limit and the FDT limit, respectively.
With this parameter scan, we construct a phase diagram that focuses on {\em active} string dynamics (Fig. \ref{fig:introductory}d-e). We measure the mean size and mean lifetime of strings, defined as time interval between changes in string {\color{black}{composition}}. With the exception of a string-less phase at high $D_R$ and $D_T$, i.e. a thermal FDT limit, the value of $D_R$ predicts string properties.  We find a phase of medium-sized strings that interact through {\em collisions} with lifetimes $\tau_l \sim 100$ when $D_R$ is relatively low ($D_R\leq 0.06$) and 
a phase of non-interacting {\em solo} strings with rapidly increasing $\tau_l$ and {\color{black}{lengths that exceed the system size and wrap the box}} when $D_R\geq0.15$. We return to this characterisation below. 

\begin{figure}[htbp]
    \centering
    \includegraphics[width=\linewidth]{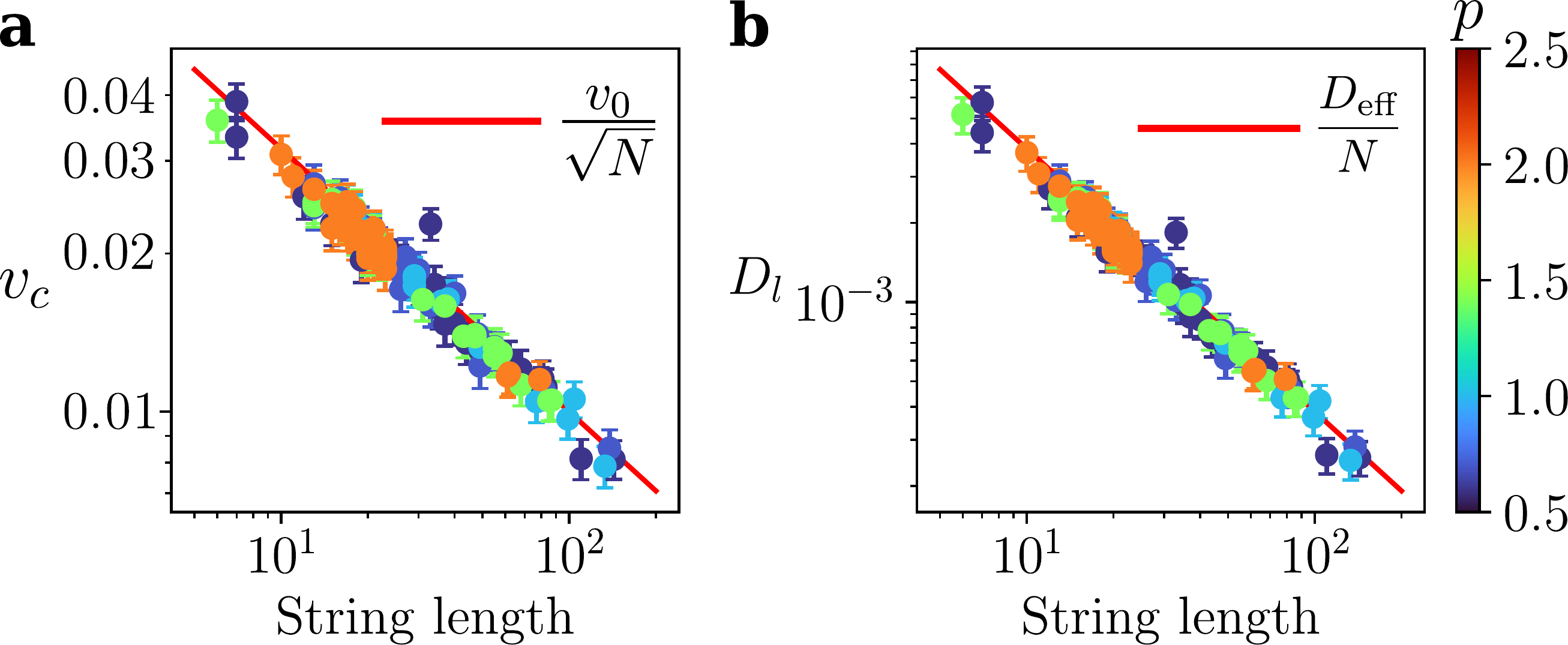}
    \caption{Collective dynamics of traveling strings as a function of their length. \textbf{a} - Self-propulsion speed of string centroids $v_c$ and  \textbf{b} - Collective {\color{black}{long-time}} translational diffusion coefficient $D_l$ (see Eqn. (\ref{collective Eqn 3})). The two red lines are our theoretical predictions Eqn. \ref{collective Eqn 3}. Data is extracted from simulations at $v_0=0.1, D_R=0.15, \frac{D_T}{\Bar{\sigma}^2D_R}=\frac{1}{30}, p\in[0.6, 0.7, 1.0, 1.4, 2.0]$, with colorbars indicating dipole strength $p$. Error bars correspond to the standard deviation of the fit. The $D_l$ data are obtained from $v_c$ and $D_c$ (Fig. S2.2); $v_c$ dominates $D_l$ in this regime resulting in similar plot shapes. Error was  propagated by assuming independent normal distributions. All data is picked from the last $5000$ time units in steady state.}
    \label{collective dynamics}
\end{figure}

We probe the collective motion of traveling strings, first focusing on the motion of their centroids. We fit the mean square displacement (MSD) of {the centroids of the strings} 
in the $xy$ plane to the MSD of a free 2-dimensional (2D) ABP in the absence of pair interactions \cite{howse2007self, breoni2020active}. The effective translational diffusion coefficient of centroids decays with string length as $D_c \sim N^{-1}$, whereas the collective self-propulsion speed decays with the square root of length $v_c \sim N^{-1/2}$ (Fig. \ref{collective dynamics}), and both are independent of dipole strength $p$, and phase. 

\textit{Active Anisotropic Rouse model ---} We can gain an understanding of string dynamics by mapping a single string to an active polymer model. By solving $\bm \nabla U_{HY}(\mathbf{r})=0$, we find the equilibrium distance between two particles $a(p)$, which is a function of the dipole strength $p$. By expanding the hybrid potential $U_{HY}$ near its stable equilibrium position $\bm {r}^{(0)}$, 
we can obtain an effective elastic potential $U_E:=\frac{1}{2}(\mathbf{r}-\bm {r}^{(0)})\cdot \mathbf{H}_U(\bm {r}^{(0)})\cdot (\mathbf{r}-\bm {r}^{(0)})^T$, with $\mathbf{H}_U$ the Hessian matrix. $\mathbf{H}_U$ is diagonal and our three effective elastic constants are given by $\kappa_{11}(p):=\frac{\partial ^2U_{HY}}{\partial x^2}\Big|_{\mathbf{r}=\bm {r}^{(0)}}=\frac{\partial ^2U_{HY}}{\partial y^2}\Big|_{\mathbf{r}=\bm {r}^{(0)}}$ and $\kappa_{33}(p):=\frac{\partial ^2U_{HY}}{\partial z^2}\Big|_{\mathbf{r}=\bm {r}^{(0)}}$. These elastic constants are functions of the dipole strength $p$ and are isotropic in the $xy$ plane. $\kappa_{11}(p)$ and $\kappa_{33}(p)$ are both strongly increasing with $p$, and $\kappa_{33}\gg \kappa_{11}$. See SI section 3.1 for details.

For a string of size $N$, the position of the $n$th particle can be expanded around a rigid column as $\bm{r}_n:=\bm{r}_{n}^{(0)} + \mathbf{R}_n$ where $\bm{r}_{n}^{(0)} = (0,0,a n)$ and fluctuations $\mathbf{R}_n = (R_{1 n},R_{2 n},R_{3 n}):=(x_n, y_n, z_n), 1\leq n\leq N$ (Fig. \ref{fig:introductory}c).
The equations of motion for the strings therefore are an active anisotropic Rouse model~\cite{doi1988theory} (see SI section 3.2),
\begin{align}
\begin{split}
    \Dot{R}_{\alpha n}&= \frac{\kappa_{\alpha \alpha} }{\zeta} \frac{\partial^2 R_{\alpha n}}{\partial n^2}+A_{\alpha n}+\eta_{\alpha n}^T, \\
    \Dot{\theta}_n&=\eta_n^R.  \label{Active Rouse model}
\end{split}
\end{align}
for $\alpha=1,2,3$. Here activity ${\bf A}_n=(A_{1n},A_{2n},A_{3n})$ is confined to the $xy$ plane, i.e $A_{1n}=v_0\cos\theta_n$, $A_{2n}=v_0\sin\theta_n$ and $A_{3n}=0$. 
Using Rouse modes~\cite{doi1988theory,Liverpool2003a}, 
the equations can be solved analytically and various collective quantities obtained. Please see SI section 3.2-3.3 for details.

The motion of the centroid  of the string is given by the lowest  (0th) Rouse mode. We therefore compute the MSD of the string centroid in the $xy$ plane to obtain
$\mbox{MSD}_{\mbox{\small string}}=4 {D_c} t+2 {v_c^2} D_R^{-1}\left[t+D_R^{-1}\left( e^{-D_R t}-1 \right) \right]$,    
with an effective translational diffusion coefficient ${D_c}$ and {the collective} self-propulsion speed ${v_c}$. Comparing the collective $\mbox{MSD}_{\mbox{\small string}}$ with that of a single ABP~\cite{howse2007self, breoni2020active}, we find  
\begin{align}
    {v_c=\frac{v_0}{\sqrt{N}}, ~\ D_c=\frac{D_T}{N}, ~\ D_l=D_c + \frac{v_c^2}{2D_R} = \frac{D_{\text{eff}}}{N}} \; .
    \label{collective Eqn 3}
\end{align}
We also {obtain $D_l$, the long-time diffusion coefficient of string centroids in terms of} $D_{\text{eff}}= D_T + v_0^2/(2 D_R)$, the long-time diffusion coefficient of an isolated Janus colloid  with both translational noise and active driving.
Our result also shows that 
the persistence time {of the strings} $D_R^{-1}$ is the same as that of a single particle. Hence, we find the collective dynamics of traveling strings is solely governed by their length and Eqn. \ref{collective Eqn 3} accurately predicts the simulation results (red lines in Fig. \ref{collective dynamics}).

We can understand the difference between {\em collisional} and {\em solo} 
strings with this simple argument: for fixed average string size $\langle N \rangle$, in the low $D_{R}$ (collisional) limit, strings move persistently for their mean free path~\cite{ChaikinLubensky95} which can be approximated as $l_f \approx \frac{\pi\langle N \rangle \Bar{\sigma}^3}{12 R L_z \phi}$, where $R$ is the $xy$ radius of gyration, {\color{black}{and $\frac{6 L_z \phi}{\pi \langle N \rangle\Bar{\sigma}^3}$ the projected $2d$ density of strings,}}  leading to an approximate lifetime $\tau_l^{\text{coll}} \approx l_f/v_c$. In contrast, in the solo regime with large $D_R$, they are moving diffusively and traverse this distance in $\tau_l^{\text{diff}} \approx \frac{l_f^2}{4 D_{l}} \gg \tau_l^{\text{coll}}$, leading to a much longer lifetime. Using $\langle N \rangle =80, L_z=80 \bar{\sigma}$ and $R={1.5}$, this is a good match for the observed string life times at low $\frac{D_T}{\Bar{\sigma}^2D_R}$ (see Fig. S{2.3}).

In addition to their persistent centroid motion, the fluctuations along the strings are also highly spatially correlated (Fig. \ref{fig:introductory}b and {SI movies}). These mesoscale spatial correlations emerge from the temporal correlations of the active driving coupling preferentially to the long wavelength elastic modes \cite{henkes2020dense,caprini2020spontaneous}. 

\begin{figure}[htbp]
    \centering
    \includegraphics[width=\linewidth]{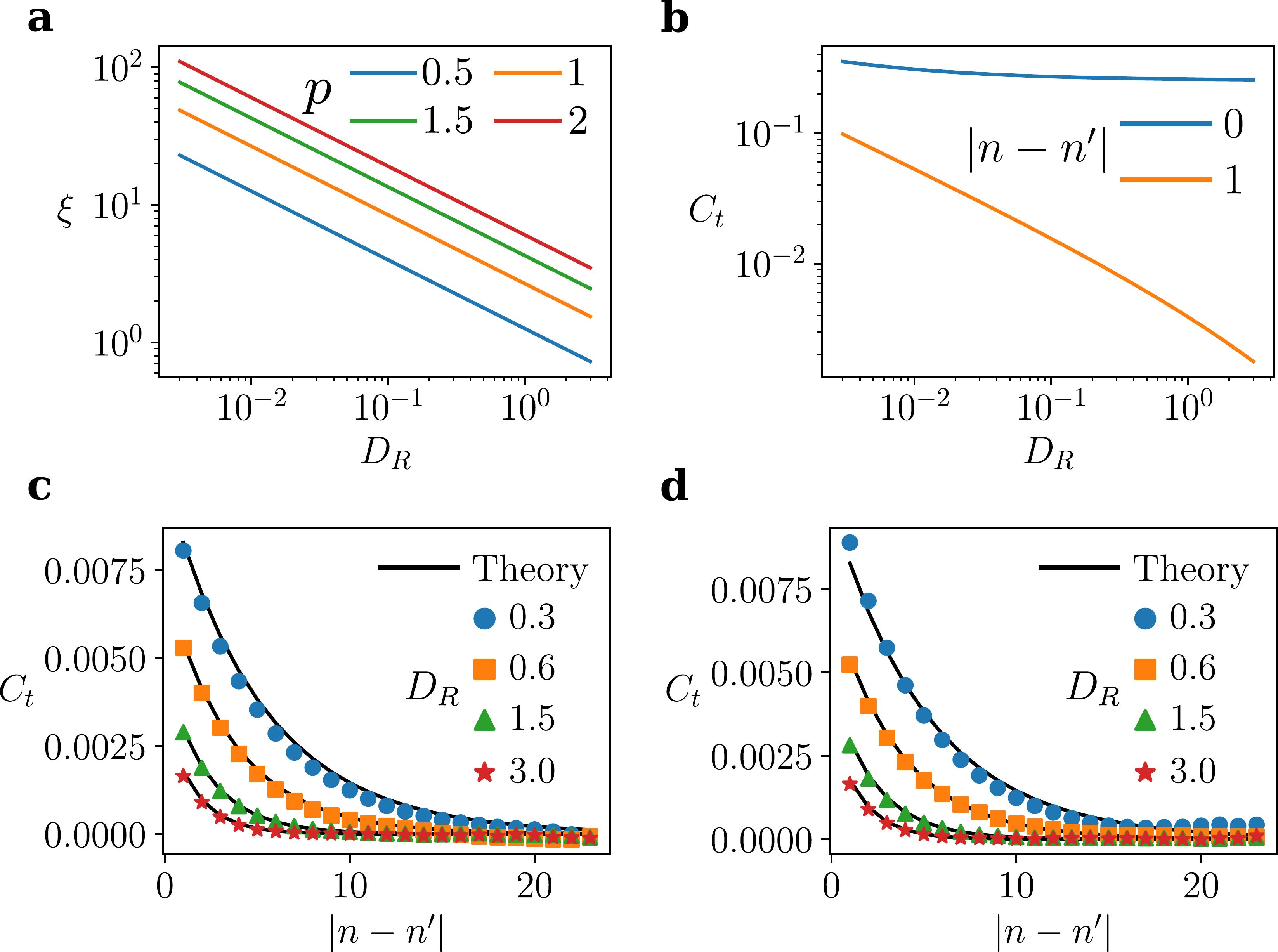}
     \caption{\textbf{a} - Correlation length $\xi\sim D_R^{-1/2}$ for different dipole strengths $p$. 
     \textbf{b} - Theoretical $C_t$ at $|n-n'|=0, 1$.  \textbf{c-d}, Bond vector correlation functions in non-interacting (\textbf{c}) and interacting (\textbf{d}) string systems. $p=1, v_0=0.5, \frac{D_T}{\Bar{\sigma}^2D_R}=\frac{1}{300}$. Points are simulation data, lines are theoretical predictions. } 
    \label{tangent correlation}
\end{figure}

To illustrate this, we analyse the correlations of the {\em bond-vectors}, i.e 
$\bm{b}_n= \bm{r}_n - \bm{r}_{n-1}$, 
at different positions on the traveling strings (see Fig. \ref{fig:introductory}c). We find new  active contributions which are due to the finite time correlations in the directions of local active driving. 
The relevant correlation function is the correlation between the {\em deviations} of the bonds from those of a rigid column: 
$\mathbf{B}_n=\mathbf{R}_n-\mathbf{R}_{n-1} = \mathbf{b}_n - (0,0,a)$.
We obtain an exact expression for this bond-vector deviation correlation function,  
$C_{t}(n,n')=\langle {\bf B}_n \cdot {\bf B}_{n'}\rangle 
$, 
using the higher Rouse modes (see SI section 3.5) :    
$C_{t}(n,n)=\frac{2aD_a}{\xi D_R}\left(e^{-\frac{a}{\xi}}-1\right)+2\zeta\left(\frac{D_T+D_a}{\kappa_{11}}+\frac{D_T}{2\kappa_{33}} \right):=C_{t0}$ when $n=n'$, and
\begin{align}
C_t (n,n')=\frac{aD_a}{\xi D_R}\left(  e^{\frac{a}{\xi}}+e^{-\frac{a}{\xi}}-2 \right)e^{-\frac{a}{\xi}|n-n'|},
\label{tan eqn 4}
\end{align}
which is valid when the bonds are far from the ends, $1\ll n\neq n'\ll N$. Here $D_a=\frac{v_0^2}{2D_R}$ is the active contribution to the ABP effective translational diffusion coefficient. The correlation length $\xi=a\sqrt{\frac{\kappa_{11}}{D_R\zeta}}$ of the exponential decay scales as $\frac{1}{\sqrt{D_R}}$, i.e. with the square root of active persistence time (Fig. \ref{tangent correlation}a). We note that in the thermal limit, the system is a flexible chain with no bond vector correlations when $n\neq n'$, i.e. the correlations are a purely active effect. Fig. \ref{tangent correlation}b shows how in the thermal limit $D_R \rightarrow \infty $, the $n=n'$ part of $C_t$ decays to a constant  proportional to $D_T$, whereas for $|n-n'|\neq 0$ it decays to $0$.
Eqn. \ref{tan eqn 4} is an excellent match to simulations of an isolated (non-interacting) string system over several $D_R$ (see Fig. \ref{tangent correlation}c). In the simulations we have subtracted the mean squared equilibrium distance between two consecutive particles, $a^2$ from $\langle \bm{b}_n \cdot \bm{b}_{n'}\rangle$. {\color{black}{For interacting strings (see Fig. \ref{tangent correlation}{d}), there is still excellent agreement if we modify the reference state for the bond deviations: $\mathbf{B}_n = \mathbf{b}_n - (0,0,a_{em})$,  with $a_{em}$ determined by an empirical least squares fit. In Fig. S2.4, we show correlations for a range of $D_R$ and the best fit $a_{em}/a$. }}

\textit{Experiment --- }
In our experiments, we use the 3D 
induced-charge electrophoresis 
system previously developed by one of us~\cite{sakai2020active}. Janus particles made from an insulating colloid half coated by metal are placed in an oscillating electric field $E$. 
We study a fixed volume fraction of $5\%$. 
In this system, the Janus colloids move like active Brownian particles in a plane orthogonal to the field ($xy$) and diffuse in the third direction ($z$). Due to the imbalance of the dielectric constant between the solvent and the particles, dipolar interactions are induced by the external electric field, which point in the direction of the field. For our parameters (5 kHz and NaCl at a concentration of $10^{-4}\mathrm{mol\cdot L^{-1}}$, i.e. $0.1$mM), 
the interactions between the particles can be approximated as a single effective dipole located at the centre. 
The Debye screening length is $\approx 26$nm. 
Here we focus on experiments with an external electric field amplitude $E={1 \over 3} \mbox{V}/\mu\mbox{m}$ 
at 5KHz frequency, giving us an individual particle P\'eclet number $\text{Pe}\approx 40$. 
Due to limited $z$ resolution, we use Trackpy to first identify particles with good $xy$ accuracy in individual $z$ layers vertically spaced by $a_{\text{eff}}=2.6\mathrm{\mu m}$, the effective particle spacing  at this salt concentration.

\begin{figure}[htbp]
    \centering
    \includegraphics[width=\linewidth]{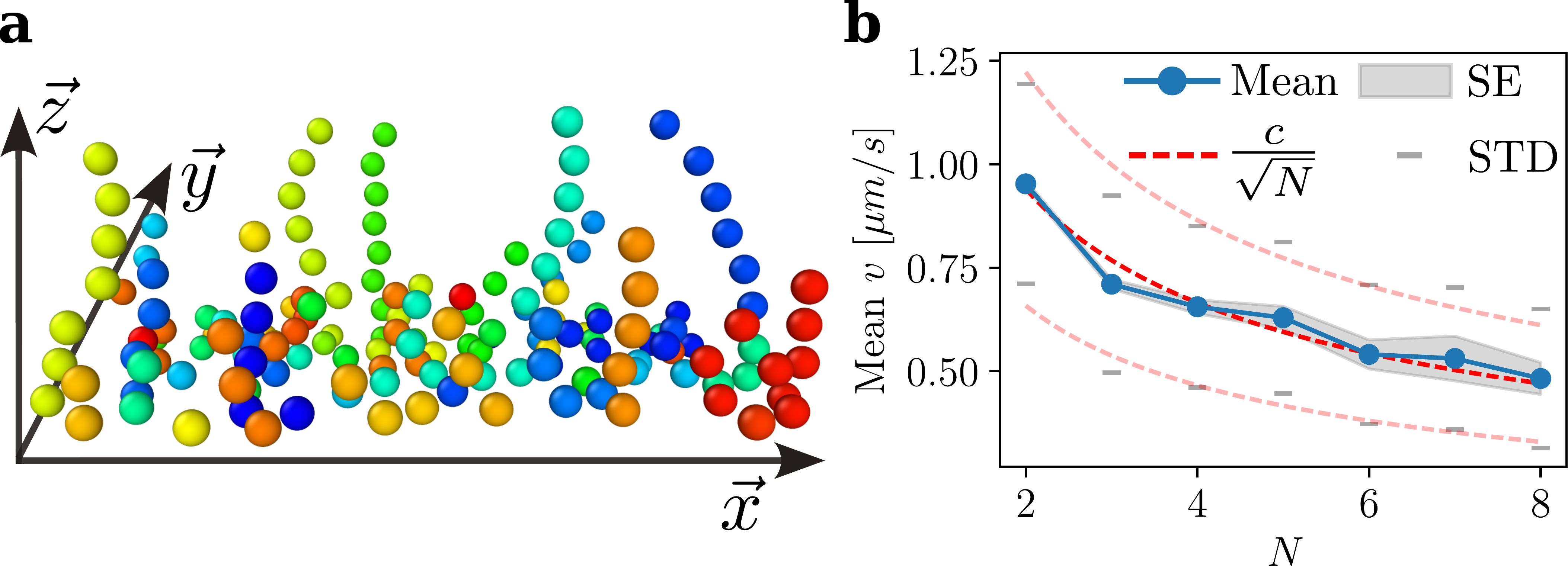}
    \caption{\textbf{a} - Coordinates from experimental data. Different colors indicate different strings. \textbf{b} - Mean instantaneous speed plot. We average the speed over strings of the same length. The best fit of the mean is $v=\frac{c}{\sqrt{N}}$ with $c=1.3$. The gray bars indicate the standard deviation and the shaded area is the standard error of the mean. } 
    \label{experiment}
\end{figure}

We find a landscape of cairn-like strings~\footnote{https://en.wikipedia.org/wiki/Cairn} (columns) with variable height growing from the bottom surface, due to the finite gravitational length (Fig. \ref{experiment}a). To identify strings, we developed a clustering algorithm that follows strings from the top to the bottom by connecting {\color{black}{to the nearest point, if any,  within $a_{\text{eff}}$.}}
%
Due to the finite {scan--time for  
each $z-$layer, moving from the bottom to the top of the image} 
{\em and} the fact that the strings are self-propelled, moving with instantaneous velocity $\mathbf{v}$ in the $xy$-plane, the observed strings appear tilted.
%
%
If the scan time of the string $ \tau_s$ is less than $D_R^{-1}$, the string approximately travels in a fixed direction during the scan (we estimate $ \tau_s \lesssim D_R^{-1}$).
We can then measure the velocity $\mathbf{v}$ of traveling strings from the tilt angle (and plane) of the strings. We perform a least-squares analysis on the beads in the string to obtain a best fit straight line, $l_s$  
making an angle $\theta_s$ with $\hat{\bf z}=(0,0,1)$.  ($\theta_s<\frac{\pi}{2}$). If in a time $\tau_s$, the camera has scanned up to height $z_s$, then the string angle $\theta_s$ and its instantaneous speed $v=|\mathbf{v}|$ are related by $v \tau_s=\tan\theta_s z_s$.
Using our {imaging} 
parameters, we find $v = 1.5802\tan\theta_s\mu \text{m}\text{s}^{-1}$ in our experiment. By averaging $v$ over strings with the same length $N$ we find that the mean $v$ decays with $N$ as $\frac{1}{\sqrt{N}}$  (Fig. \ref{experiment}b),
which is consistent with our simulations and theory.


In conclusion, we have studied a string forming 3D active dipolar colloidal system using simulations, theory and experiment. The collective dynamics of traveling strings, derived analytically and confirmed by numerical simulations and experimental analyses, has a simple dependence on string length. At low packing fractions, the string dynamics is well described by an active anisotropic Rouse model, showing emergent active bond vector correlations. 
In future work we plan to extend our analysis to the active sheets and labyrinth appearing at higher packing fractions~\cite{sakai2020active}, where we also expect additional hydrodynamic contributions to string dynamics. 


\begin{acknowledgments}
The authors would like to thank the Isaac Newton Institute for Mathematical Sciences, Cambridge, for support and hospitality during the programme {\em New statistical physics in living matter}, where part of the work on this paper was undertaken. This work was supported by EPSRC grants EP/R014604/1 and EP/T031077/1. X.C. is supported by Bristol-CSC joint program.
\end{acknowledgments}

\bibliography{ref_paper}

\end{document}